\documentclass[%
 reprint,
 superscriptaddress,
 amsmath,amssymb,
 aps,
 prl,
]{revtex4-2}

\usepackage{graphicx}
\usepackage{dcolumn}
\usepackage{commath}
\usepackage{bm}
\usepackage{comment}
\usepackage[colorlinks=true,linkcolor=blue]{hyperref}
\hypersetup{
    colorlinks=true,
    linkcolor=cyan,
    filecolor=magenta,      
    urlcolor=blue,
    citecolor=blue,
}

\usepackage{soul}
\usepackage{lipsum}
\usepackage{xcolor}

\begin{document}

\preprint{APS/123-QED}

\title{
Flowing menisci: coupled dynamics and liquid exchange with soap films
}


\author{Alexandre~Vigna-Brummer}
\email{alexandre.vigna-brummer@univ-cotedazur.fr}
\affiliation{Universit\'e C\^ote d'Azur, CNRS, INPHYNI, Nice, France}%

\author{Antoine~Monier}
\affiliation{Universit\'e C\^ote d'Azur, CNRS, INPHYNI, Nice, France}%

\author{Isabelle~Cantat}
\affiliation{Universit\'e de Rennes, CNRS, IPR (Institut de Physique de Rennes) - UMR 6251, 35000 Rennes, France}%
\affiliation{Institut Universitaire de France (IUF), Paris, France}

\author{Christophe~Brouzet}
\affiliation{Universit\'e C\^ote d'Azur, CNRS, INPHYNI, Nice, France}%

\author{Christophe~Raufaste}
\email{christophe.raufaste@univ-cotedazur.fr}
\affiliation{Universit\'e C\^ote d'Azur, CNRS, INPHYNI, Nice, France}%
\affiliation{Institut Universitaire de France (IUF), Paris, France}%

\date{\today}

\begin{abstract}
Liquid foams exhibit menisci whose lengths range from hundreds of microns in microfoams to several centimeters in macroscopic bubble arrangements. These menisci thin under gravity until reaching a steady thickness profile, where hydrostatic and capillary pressures are balanced.
While these menisci are in contact with soap films, dynamical liquid exchanges between them are neglected in current drainage models, which assume thin films provide a negligible liquid reservoir. 
Using controlled experiments, we systematically measure the shape of an isolated meniscus placed in a vertical soap film. By increasing the film thickness, we identify a flowing regime in which the flux from the adjacent film significantly enlarges the meniscus. We present an analytical model that extends the drainage equation to incorporate this film flux and introduce a gravito-exchange length, which sets the minimum meniscus thickness. The model is in quantitative agreement with experiments, capturing the transition between hydrostatic and flowing menisci. This study has implications for flowing or rearranging foams, where thick films are commonly observed. 
\end{abstract}

\maketitle

\emph{Introduction} - Liquid foams, dense dispersions of gas bubbles in a liquid phase, have a unique biphasic nature that enables a wide range of applications \cite{cantat2013foams}. These include uses in the food, cosmetics, and pharmaceutical industries, as well as in energy and environmental sectors for mineral extraction, gas recovery, soil remediation, CO$_2$ sequestration, and geothermal energy exploitation~\cite{stevenson_foam_2012,sun_research_2023,chaudhry_recent_2024,harshini_innovative_2024}. 
The stability and rheological properties of a foam strongly depend 
on the distribution of the liquid phase within the films separating neighboring bubbles and throughout the interconnected network of menisci, called Plateau borders  \cite{cantat2013foams}. This distribution is highly dynamic, with constant liquid exchange between films and menisci. Foams have evolving structures where bubble rearrangements, known as T1 events \cite{weaire}, continuously occur, leading to the formation of new thick films with initial thicknesses that can reach several micrometers \cite{petit2015generation}. The films undergo cycles of formation during T1 events and thinning by drainage, with their average thickness in a foam sample determined by the relative rates of these processes \cite{saint-jalmes_foam_2023}. The frequency of T1 events is driven by foam coarsening \cite{cohenaddad01} or the applied shear rate \cite{gopal95}, while the rate of film drainage is governed by capillary suction toward the menisci \cite{chatzigiannakis_thin_2021}. 

This liquid exchange is driven by a difference in Laplace pressure, $\gamma/r$, between the flat film of thickness~$h$ and a concave meniscus of radius of curvature~$r$, where $\gamma$ is the surface tension. 
This pressure difference induces a flux of liquid, from the film to the meniscus, per unit length of meniscus, which scales as $\gamma h^{5/2}/(\eta r^{3/2})$~\cite{mysels1959soap, lhuissier2012bursting, gros2021marginal} (see End Matter), where $\eta$ is the dynamic viscosity. 
Former studies addressing this mechanism of liquid exchange assumed a constant, imposed, value for $r$, whereas it is a dynamic quantity that varies with both space and time. 
At equilibrium, the vertical menisci are expected to display inhomogeneous shapes determined by the balance between Laplace pressure and gravity, $\vec{0} = \vec{\nabla}(\gamma/r) + \rho_l \vec{g}$, where $\rho_l$ is the liquid density and $\vec{g}$ is the gravitational acceleration. This balance gives an expression for the radius of curvature as a function of height, 
\begin{equation}\label{Eq:RmCapillary}
r(z) = \lambda_c^2/(z-z_{\infty}) , 
\end{equation}
where $\lambda_c = \sqrt{\gamma/(\rho_l g)}$ is the capillary length and  $z_\infty$ a constant that depends on boundary conditions. 
When displaced from equilibrium, menisci relax by redistributing liquid at their own scale \cite{cohen2014inertial, argentina2015one, cohen2015drop}, and through the entire network of interconnected menisci, as described by the drainage equation \cite{koehler_generalized_2000, saint-jalmes_physical_2006}.

Traditional models assume that films stay near their equilibrium thickness and do not affect drainage in the menisci. In this study, we show that for sufficiently thick films, the liquid flux from the film can significantly alter the meniscus shape, necessitating a coupled analysis of fluid motion in both the film and menisci. 
To highlight this coupling, 
we examined the steady shape of a single and isolated meniscus formed at the contact between a vertical soap film and a ring. The spatial variation of the meniscus radius was accurately predicted by an extended drainage equation that incorporates the flux from the film. 
Our findings reveal a new characteristic length scale, the gravito-exchange length, which determines the minimum meniscus thickness. 

\begin{figure*}[t!]
\centering
\includegraphics[width=\textwidth]{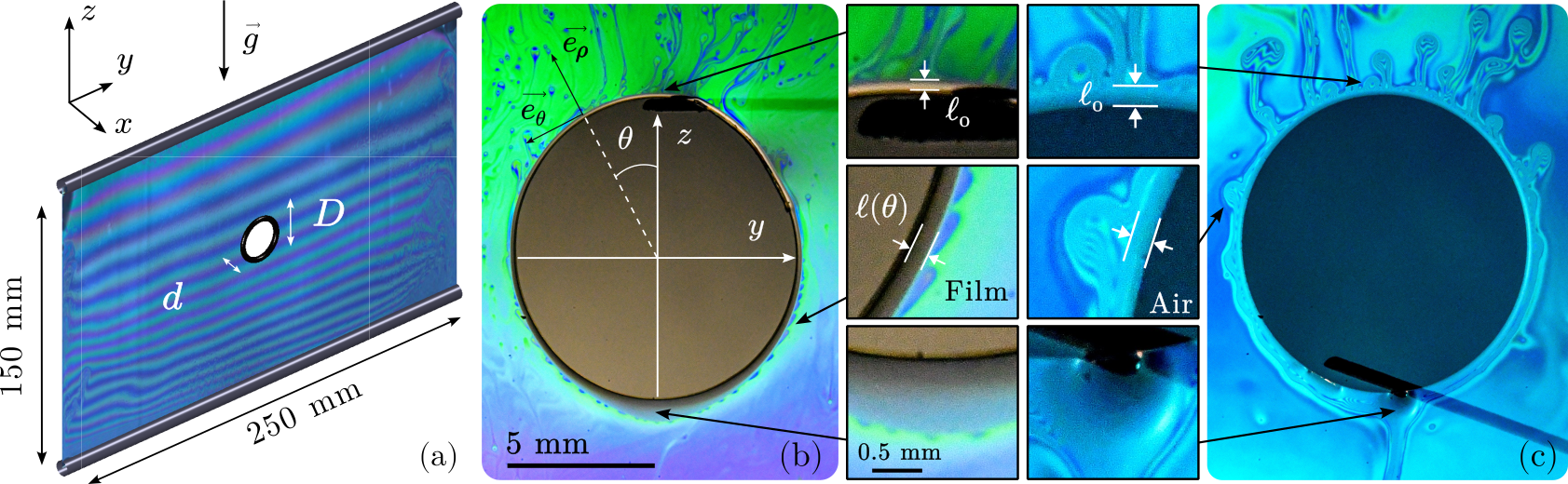}
\caption{
(a)~Setup: a ring (with dimensions $D$ and $d$) is suspended in a soap film with a controlled thickness profile and held in a fixed position by a rod.  
(b) Image with $D = 9.8$~mm, $d=52~\mu$m and film thickness $h=0.4~\mu$m at the top of the ring. Magnified views from the top, side and bottom of the rings are shown.
The soap film, appearing colorful due to light interference in reflection, contrasts with the dark appearance of the ring and the gray meniscus of extent $\ell$.
(c) Same experiment with a smaller minor diameter and a thicker film $(d=14.5~\mu$m and $h=7.5~\mu$m). Here the ring is not distinguishable while the meniscus is also colorful and extends up to the base of the rising thin film elements. Typical movies in Supplementary~Material~\cite{SuppMat}.
}
\label{Fig:Notations}
\end{figure*}

\emph{Materials and Methods} - Vertical soap films were created within a rectangular frame measuring $150$~mm in height and $250$~mm in width. The frame consists of a perforated pipe at the top, connected to a rod at the bottom by two vertical strings, as detailed in Fig.~\ref{Fig:Notations}(a). A surfactant solution is injected into the perforated pipe to supply the film and counteracts gravitational drainage, thus leading to a stationary thickness profile \cite{adami2014capillary}. A constant flow rate between $0.6$ and $1.2$~mL.min$^{-1}$ was maintained using a syringe pump.
Three different surfactant solutions were tested to confirm the universality of the results (see End Matter). 

Ring torus, of major and minor diameters noted~$D$ and~$d$, were introduced into the vertical soap film, as shown in Fig.~\ref{Fig:Notations}. The rings tend to rise or sink within the film due to capillary-driven buoyancy \cite{plateau1873, couder1989hydrodynamics, adami2014capillary}. To counteract this motion, they are held in a fixed position using one or two rods that contact the ring internally. 
Four types of materials are used to construct the rings, allowing for control over the minor diameter~$d$: glass fibers ($d=14.5\pm0.5~\mu$m), human hairs ($d=52 \pm 2~\mu$m), nylon ($d=334\pm 2~\mu$m) and fluorocarbon ($d=609\pm 10~\mu$m) fibers. The major diameter of these rings ranges from $D=5.6$ to $65.4$~mm. 

Varying the flow rate and the type of surfactant modifies the thickness profile of the soap film, as measured by thin-film interferometry (see End Matter). Additionally, adjusting the ring's position within the film further extends the range of thicknesses in the film portion in contact with the ring. The thinnest films were achieved by stopping the syringe pump, allowing the film to drain under gravity. Overall, the thickness~$h$ of the film surrounding the ring varied between $0.4$ and $10~\mu$m, with a maximum difference of $30\%$ between the top and bottom of the ring.

Figure~\ref{Fig:Notations}(b) shows a typical image and close-ups of a ring introduced into a soap film, while Fig.~\ref{Fig:Notations}(c) presents a case with a similar major diameter $D$ but a smaller minor diameter $d$ and a larger film thickness $h$. 
In both cases, a meniscus is visible at the contact between the ring and the soap film. Its extent $\ell$ is measured as the distance between the outer part of the ring and the base of the rising thin film elements described below. The meniscus is narrowest at the top, where it has a finite extent $\ell_0$, and widens toward the bottom. Although the meniscus shapes are steady, liquid exchange between the meniscus and the film is evident. In the case of Fig.~\ref{Fig:Notations}(b), this exchange is marked by occasional dripping of meniscus elements at the bottom (see movie in Supplementary Material \cite{SuppMat}) and by the phenomenon of marginal regeneration, characterized by thin film elements nucleating, rising along the meniscus, and escaping into the soap film at the top of the ring \cite{mysels1959soap, lhuissier2012bursting, gros2021marginal}.
In Fig.~\ref{Fig:Notations}(c), liquid exchange is even more pronounced, as evidenced by the continuous evacuation of liquid through a channel at the bottom and by the increased activity of thin film elements. 

\begin{figure}[t!]
\centering
\includegraphics[width=0.45\textwidth]{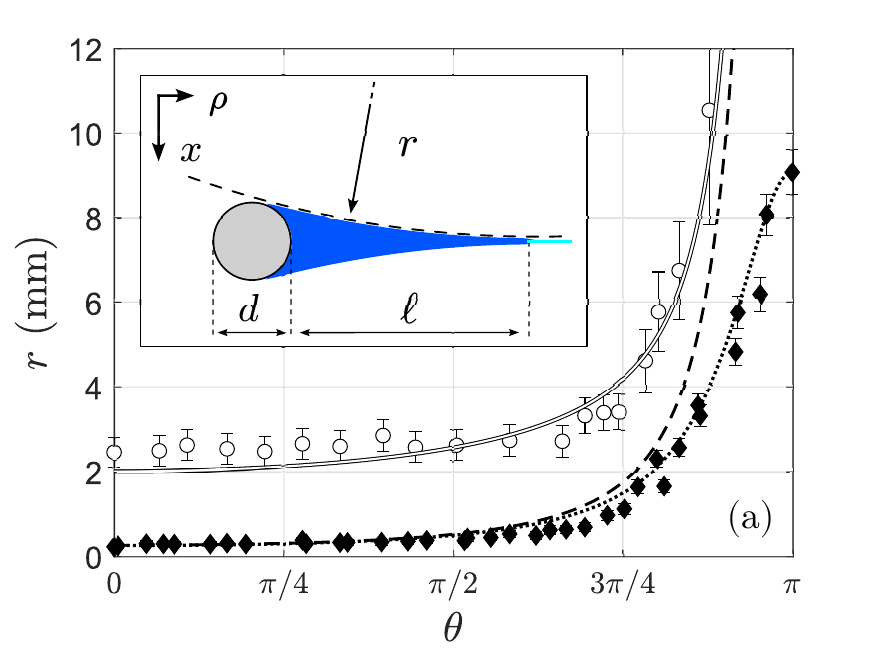} 
\includegraphics[width=0.45\textwidth]{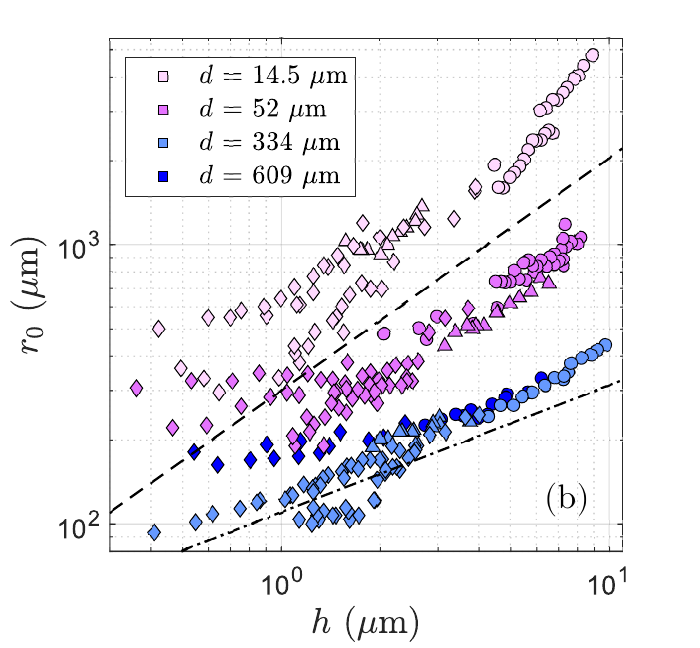}
\caption{\label{Fig:Result1}
(a) Profiles $r(\theta)$.  
Black diamonds: same series as in Fig.~\ref{Fig:Notations}(b). With $r_\pi$ found experimentally, the solutions of Eq.~\eqref{eq:static_ring} is indistinguishable from the solution of Eq.~\eqref{Eq:dimensionequation_all} with $r_g/r_s=0.22$,  and is represented by the dotted line.
White circles: same series as in Fig.~\ref{Fig:Notations}(c), with the dashed line solution of Eq.~\eqref{eq:static_ring} with $r_\pi \rightarrow +\infty$.  
The solid line is the solution of 
Eq.~\eqref{Eq:dimensionequation_all}, with $r_g/r_s=8$. 
Inset shows the meniscus in blue, the ring in gray and the soap film in cyan, in a ($\rho$, $x$) cross-section as defined in Fig.~\ref{Fig:Notations}(b). (b) $r_0$ as a function of film thickness $h$ for all experiments, including various surfactant solutions (Dreft with diamond, SDS with triangle and SLES+CAPB with circle), and the full ranges of ring major and minor diameters, $D$ and $d$. 
The two lines show power-laws with exponent given by the model: $5/6$ (dashed) and $5/11$ (dashed-dotted).
}
\end{figure}

\emph{Results} - 
In the images, we measured the meniscus extent $\ell$ as a function of the angular position $\theta$, defined in Fig.~\ref{Fig:Notations}(b), with $\theta$ varying between $0$ and $\pi$. 
We inferred the meniscus radius of curvature~$r$ from~$\ell$, given their geometric relationship with the ring's minor diameter~$d$: $r=\ell(1+\ell/d)$ (see inset of Fig.~\ref{Fig:Result1}(a) and End Matter). In Fig.~\ref{Fig:Result1}(a), $r$ is plotted as a function of $\theta$ for the two cases shown in Figs.~\ref{Fig:Notations}(b) and~(c). For the first case, the data agree well with a hydrostatic meniscus model (dotted curve in Fig.~\ref{Fig:Result1}(a)), where $z=\frac{D}{2}(1+\cos\theta)$ is used in Eq.~\eqref{Eq:RmCapillary}, and $z_\infty$ is adjusted to match the radius $r(\theta=\pi)=r_\pi$ at the bottom of the ring, 
\begin{equation}
r(\theta)=\frac{r_s}{\frac{1+\cos \theta}{2}+\frac{r_s}{r_\pi}},\label{eq:static_ring}
\end{equation}
with $r_s=\lambda_c^2/D$. Since $r_s\ll r_\pi$, we can approximate $r_s$ as the radius of curvature of the meniscus at the top of the ring ($\theta=0$). Importantly, $r_s$ is independent of film thickness~$h$ and ring's minor diameter~$d$, depending only on the ring's major diameter~$D$ and capillary length~$\lambda_c$. In the second case, where liquid flow is significant, the radius of curvature 
diverges as $\theta$ approaches~$\pi$, i.e. $r_\pi \rightarrow +\infty$. This would yield the meniscus radius of curvature as $r(\theta)=2 r_s/(1+\cos \theta)$ (dashed curve in Fig.~\ref{Fig:Result1}(a)). However, this hydrostatic meniscus model cannot predict the data in that case, since the radius of curvature is typically an order of magnitude larger than~$r_s$ in the region between $0$ and $\pi/2$.

We therefore focused on the top of the ring, by measuring $r_0=r(\theta=0)$. We found that $r_0$ is an increasing function of the film thickness $h$ (Fig.~\ref{Fig:Result1}(b)): at large~$h$, we observed power-laws highlighted by the dashed and dashed dotted lines, while at small~$h$, a saturation is visible for some of the data. $r_0$ is also significantly affected by the minor ring diameter~$d$, but seems to be less dependent on the ring major diameter~$D$, as illustrated by the limited dispersion of the points within the same~$d$.

\emph{Model} - To explain these experimental results, we here develop a model. Inside the meniscus, the liquid flows along the azimuthal direction. In cylindrical coordinates~$(\rho,\theta,x)$, as defined in Fig.~\ref{Fig:Notations}(b), the velocity field can be expressed as $\vec{v} = v \, \vec{e}_\theta$. 
In the Stokes equation, we consider only the dominant terms in the pressure and viscous contributions under the assumption that the cross-sectional dimensions of the meniscus $(\ell$ and~$d$, inset of Fig.~\ref{Fig:Result1}(a)) are significantly smaller than the ring major diameter~$D$. This leads to
\begin{equation}\label{eq:NS}
    0 = -\frac{2}{D}\frac{\partial P}{\partial\theta} + \rho_l g\sin\theta + \eta \left(   \frac{\partial^2 v}{\partial \rho^2} + \frac{\partial^2 v}{\partial x^2} \right),
\end{equation}
where $P$ is the pressure inside the meniscus.  
According to Laplace's law, the pressure $P$ is given by $P = P_0 - \gamma/r$, {with $P_0$ the ambient pressure}. 
Given the no-slip boundary condition at the contact with the ring, we anticipate larger velocity gradients along the radial direction. Consequently, the average speed $\bar{v}$ in the meniscus can be reasonably estimated from Eq.~\eqref{eq:NS} by the following relation, 
\begin{equation}\label{Eq:v_simplified}
\bar{v} \sim   \frac{\gamma}{\eta} \ell^2 \left(  -\frac{2}{D r^2} \frac{\partial r}{\partial\theta} + \frac{\sin\theta}{\lambda_c^2} \right) .
\end{equation}

Given the concavity of the meniscus (inset of Fig.~\ref{Fig:Result1}(a)), the pressure within the meniscus is lower than that in the soap film, driving liquid exchange from the film to the meniscus~\cite{mysels1959soap, lhuissier2012bursting, gros2021marginal}.
In the steady state, this liquid influx balances the azimuthal variation of the mass flux inside the meniscus, such as
\begin{equation}\label{Eq:fluxbalance}
\frac{\mathrm{d} (A \bar{v})}{\mathrm{d}\theta}=\frac{qD}{2} ,
\end{equation}
where $A$ is the cross-sectional area of the meniscus and $q\sim\gamma h^{5/2}/(\eta r^{3/2})$ the liquid flux per unit length of meniscus (see End Matter).

\begin{figure}[t!]
\centering
\includegraphics[width=0.45\textwidth]{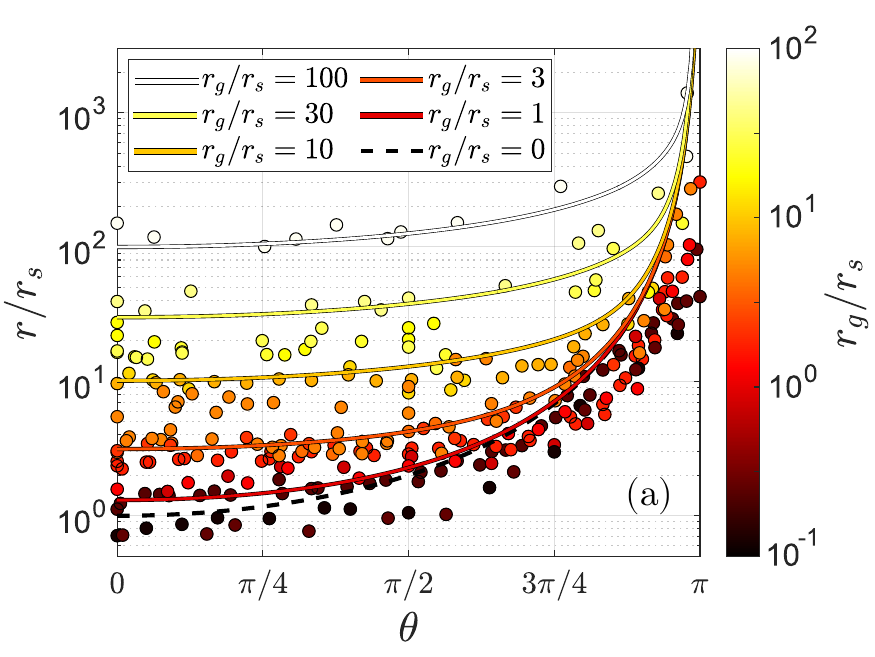}
\includegraphics[width=0.45\textwidth]{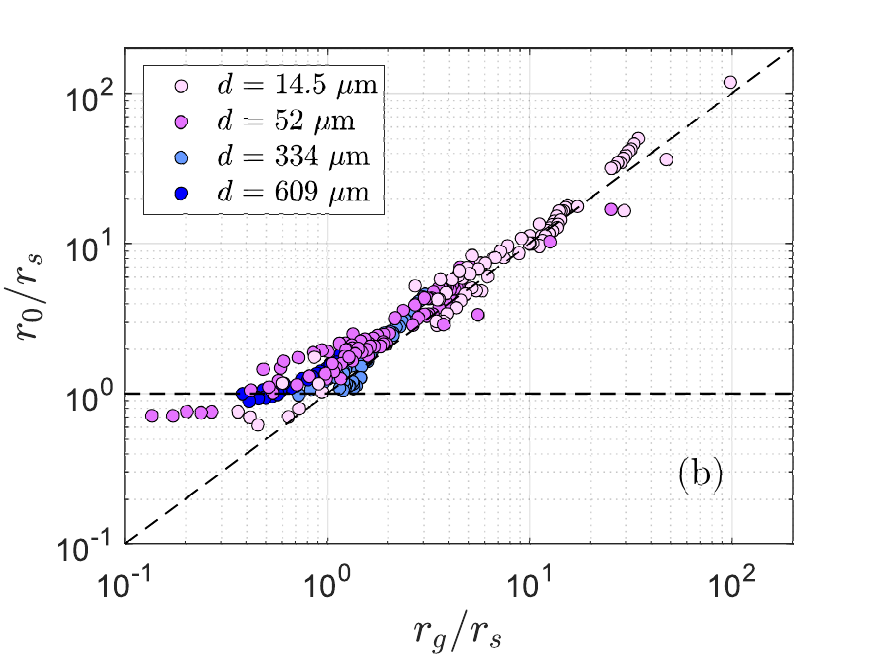}
\caption{\label{Fig:Model1}
(a) $r/r_s$ as a function of $\theta$ for various ring geometries and film thicknesses in semi-log scale. The colorbar in log-scale indicates the corresponding~$r_g/r_s$. The dashed line is the analytical solution in the hydrostatic limit, Eq.~\eqref{eq:static_ring}, while the solid lines are the numerical solutions for different~$r_g/r_s$ of Eq.~\eqref{Eq:dimensionlessequation_all}, obtained in both cases by enforcing a divergence of $r$ at $\theta=\pi$. (b)~$r_0/r_s$ as a function of $r_g/r_s$ in log-log scale for all data of Fig.~\ref{Fig:Result1}(b). The dashed lines show the hydrostatic ($r_0=r_s$, $r_g/r_s \ll 1$) and flowing~($r_0=~r_g$,~$r_g/r_s~\gg~1$)~regimes.
}
\end{figure}

Setting $\bar{v}=0$ in Eq.~\eqref{Eq:v_simplified}  recovers the hydrostatic solution given by Eq.~\eqref{eq:static_ring}. However, we now consider the flowing case where $\bar{v}\neq 0$. Since $A$ and $r$ are non-trivially related to $\ell$ and $d$ (see End Matter), we need to make some approximations.  Two limiting cases can be identified depending on the ratio $d/r$. We here consider the small minor diameter limit, and perform the calculations for the other limit in the End Matter. 
When $d\ll r$, the relations simplify to $r \simeq \ell^2 /d$ and $A \simeq r^{1/2} \, d^{3/2}/3$, respectively. From Eqs.~\eqref{Eq:v_simplified} and 
\eqref{Eq:fluxbalance}, we get
\begin{equation}\label{Eq:dimensionequation_all}
    r^{3/2} \frac{\mathrm{d}}{\mathrm{d}\theta}\left(r^{3/2} \sin\theta - \frac{2 \lambda_c^2}{D r^{1/2}}\frac{\mathrm{d}r}{\mathrm{d}\theta}\right) = k \lambda_c^{2}D\, d^{-5/2}h^{5/2},
\end{equation}
where $k$ is a constant to account for diverse proportionality constants discarded above, expected of order~$1$. Introducing the gravito-exchange radius of curvature $r_{g} = k^{1/3} \lambda_c^{2/3}D^{1/3}d^{-5/6}h^{5/6}$, we can write Eq.~\eqref{Eq:dimensionequation_all} with the dimensionless quantity~$\tilde{r} = r/r_s$, 
\begin{equation}\label{Eq:dimensionlessequation_all}
    \tilde{r}^{3/2} \frac{\mathrm{d}}{\mathrm{d}\theta}\left(\tilde{r}^{3/2} \sin\theta - \frac{2}{\tilde{r}^{1/2}} \frac{\mathrm{d}\tilde{r}}{\mathrm{d}\theta}\right) = (r_g/r_s)^3 ,
\end{equation}
with 
$r_g/r_s=k^{1/3}(D/\lambda_c)^{4/3}(h/d)^{5/6}$. 
For a given~$r_g/r_s$, this equation is solved numerically with the boundary conditions $\textrm{d}\tilde{r}/\textrm{d}\theta$ = 0 at $\theta=0$, ensuring symmetry, and $\tilde{r}(\pi) \rightarrow +\infty$, representing the evacuation of liquid at the bottom of the ring in steady state. 
In experiments, we measured $k=1.8\pm0.1$ (see End Matter). Using this value, $r_g/r_s$ spans from $0.1$ to $100$ across our set of parameters. 
Two asymptotic behaviors for $\tilde{r}$ can be identified. 
For $r_g/r_s \ll 1$, the flux term, represented by the constant $(r_g/r_s)^3$ on the right-hand side of Eq.~\eqref{Eq:dimensionlessequation_all}, is negligible for all values of $\theta$. In this case, Eq.~\eqref{Eq:dimensionlessequation_all} reduces to the hydrostatic equation, with an analytical solution given in Eq.~\eqref{eq:static_ring}. At the top of the ring, we have $\tilde{r}(0) \approx 1$,  or equivalently, $r_0 \approx r_s$. 
For $r_g/r_s \gg 1$, the flux term remains negligible at the bottom of the ring ($\theta \sim \pi$), where variations in $r(\theta)$ are strongest. However, it becomes dominant over the capillary term (the second term on the left-hand side of Eq.~\eqref{Eq:dimensionlessequation_all}) once $\tilde{r} \sim r_g/r_s$. Higher up in the ring, the radius remains nearly constant, as the flux term is balanced by gravity (the first term on the left-hand side of Eq.~\eqref{Eq:dimensionlessequation_all}), leading to $\tilde{r}(0) \approx r_g/r_s$, or equivalently, $r_0 \approx r_g$. 

By normalizing all the experimental measurements of~$r$ by the hydrostatic limit~$r_s=\lambda_c^2/D$, we recover these features in Fig.~\ref{Fig:Model1}. The experimental and numerical profiles $r(\theta)/r_s$ are shown in Fig.~\ref{Fig:Model1}(a), with color coding for~$r_g/r_s$. At large~$\theta$, all data collapse to a master curve, following the hydrostatic curve (dashed line). At small~$\theta$, the data are classified by~$r_g/r_s$: for $r_g/r_s \leq 1$, they collapse to the hydrostatic case while for larger~$r_g/r_s$, they start from higher and higher values, as $r_g/r_s$ increases.
The larger~$r_g/r_s$, the larger~$\theta$ to recover the hydrostatic profile.
In all these comparisons, the agreement between the experiments and the numerical solutions of Eq.~\eqref{Eq:dimensionlessequation_all} is good. In particular, the numerical solutions 
for the two example profiles corresponding to $r_g/r_s\approx 0.22$ and~$8$ shown in Fig.~\ref{Fig:Result1}(a) match with the experimental data. 

In Fig.~\ref{Fig:Model1}(b), we confirmed that $r_0$ is switching from $r_s$ to~$r_g$, when $r_g/r_s$ increases. 
This collapses all the data shown in Fig.~\ref{Fig:Result1}(b) into a single master curve, where the transition between hydrostatic and flowing regimes is clearly visible, around $r_g/r_s\approx 1$. In the flowing regime, $r_0$ equals $r_g$ and is expected to scale as $(h/d)^{5/6}$ for the small minor diameter case, and as $h^{5/11}$ for the large minor diameter case (see End Matter). This thus explains the trend of the data in Fig.~\ref{Fig:Result1}(b), showing a separation with respect to the minor diameter~$d$ for small~$d$ only, and two different scalings with film thickness for small and large~$d$. 

\emph{Discussion} - This study demonstrates the impact of capillary suction and liquid influx from the films on the shape of menisci. We extend the traditional drainage equation for menisci by adding a source term to account for this effect. The influence is characterized by the gravito-exchange length, $r_g$, which determines the radius of curvature of a meniscus when the effect of liquid exchange dominates over the capillary pressure gradient. 
The transition between hydrostatic and flowing meniscus regimes is governed by the dimensionless parameter~$r_g / r_s$. For \( r_g/r_s < 1 \), the meniscus follows the hydrostatic regime, with a radius of curvature that decreases with height. For \( r_g/r_s > 1 \), the lower part of the meniscus remains hydrostatic, while its upper part saturates at~\( r_g \). As \( r_g/r_s \) increases, a larger portion of the meniscus adopts this saturated radius.  

Although this study focuses on menisci formed at the contact with circular objects,  the results can be generalized to other geometries. 
For example, the behavior of a meniscus at the contact between a film and a vertical plate, or a Plateau border at the intersection of three bubbles, can be derived by locally projecting the Stokes equation along the direction of the meniscus, and by setting the limit \( d \gg r \). The scaling is thus $r_g \sim \lambda_c^{4/11} D^{2/11} h^{5/11}$, where $D$ now represents the object's typical size or foam height. In these cases, the meniscus radius of curvature saturates to $r_g$ when the capillary radius of curvature ($r_s=\lambda_c^2/D$ from Eq.~\eqref{Eq:RmCapillary}) falls below this value. For a $1~\mu$m-thick film, the transition occurs around {a few centimeters}  for foam height or object size, with a meniscus radius of curvature of approximately $100~\mu$m.

\vspace{0.25cm}

This work was supported by Agence Nationale de la Recherche (ANR-20-CE30-0019). The authors are grateful to Emmanuelle Rio and Alice Etienne-Simonetti for fruitful discussions.

\bibliography{Biblio_universalprofile}

\begin{thebibliography}{25}%
\makeatletter
\providecommand \@ifxundefined [1]{%
 \@ifx{#1\undefined}
}%
\providecommand \@ifnum [1]{%
 \ifnum #1\expandafter \@firstoftwo
 \else \expandafter \@secondoftwo
 \fi
}%
\providecommand \@ifx [1]{%
 \ifx #1\expandafter \@firstoftwo
 \else \expandafter \@secondoftwo
 \fi
}%
\providecommand \natexlab [1]{#1}%
\providecommand \enquote  [1]{``#1''}%
\providecommand \bibnamefont  [1]{#1}%
\providecommand \bibfnamefont [1]{#1}%
\providecommand \citenamefont [1]{#1}%
\providecommand \href@noop [0]{\@secondoftwo}%
\providecommand \href [0]{\begingroup \@sanitize@url \@href}%
\providecommand \@href[1]{\@@startlink{#1}\@@href}%
\providecommand \@@href[1]{\endgroup#1\@@endlink}%
\providecommand \@sanitize@url [0]{\catcode `\\12\catcode `\$12\catcode
  `\&12\catcode `\#12\catcode `\^12\catcode `\_12\catcode `\%12\relax}%
\providecommand \@@startlink[1]{}%
\providecommand \@@endlink[0]{}%
\providecommand \url  [0]{\begingroup\@sanitize@url \@url }%
\providecommand \@url [1]{\endgroup\@href {#1}{\urlprefix }}%
\providecommand \urlprefix  [0]{URL }%
\providecommand \Eprint [0]{\href }%
\providecommand \doibase [0]{http://dx.doi.org/}%
\providecommand \selectlanguage [0]{\@gobble}%
\providecommand \bibinfo  [0]{\@secondoftwo}%
\providecommand \bibfield  [0]{\@secondoftwo}%
\providecommand \translation [1]{[#1]}%
\providecommand \BibitemOpen [0]{}%
\providecommand \bibitemStop [0]{}%
\providecommand \bibitemNoStop [0]{.\EOS\space}%
\providecommand \EOS [0]{\spacefactor3000\relax}%
\providecommand \BibitemShut  [1]{\csname bibitem#1\endcsname}%
\let\auto@bib@innerbib\@empty
\bibitem [{\citenamefont {Cantat}\ \emph {et~al.}(2013)\citenamefont {Cantat},
  \citenamefont {Cohen-Addad}, \citenamefont {Elias}, \citenamefont {Graner},
  \citenamefont {H{\"o}hler}, \citenamefont {Pitois}, \citenamefont {Rouyer},\
  and\ \citenamefont {Saint-Jalmes}}]{cantat2013foams}%
  \BibitemOpen
  \bibfield  {author} {\bibinfo {author} {\bibfnamefont {I.}~\bibnamefont
  {Cantat}}, \bibinfo {author} {\bibfnamefont {S.}~\bibnamefont {Cohen-Addad}},
  \bibinfo {author} {\bibfnamefont {F.}~\bibnamefont {Elias}}, \bibinfo
  {author} {\bibfnamefont {F.}~\bibnamefont {Graner}}, \bibinfo {author}
  {\bibfnamefont {R.}~\bibnamefont {H{\"o}hler}}, \bibinfo {author}
  {\bibfnamefont {O.}~\bibnamefont {Pitois}}, \bibinfo {author} {\bibfnamefont
  {F.}~\bibnamefont {Rouyer}}, \ and\ \bibinfo {author} {\bibfnamefont
  {A.}~\bibnamefont {Saint-Jalmes}},\ }\href@noop {} {\emph {\bibinfo {title}
  {Foams: structure and dynamics}}}\ (\bibinfo  {publisher} {OUP Oxford},\
  \bibinfo {address} {Oxford},\ \bibinfo {year} {2013})\BibitemShut {NoStop}%
\bibitem [{\citenamefont {Stevenson}(2012)}]{stevenson_foam_2012}%
  \BibitemOpen
  \bibfield  {author} {\bibinfo {author} {\bibfnamefont {P.}~\bibnamefont
  {Stevenson}},\ }\href
  {https://onlinelibrary.wiley.com/doi/abs/10.1002/9781119954620} {\emph
  {\bibinfo {title} {Foam {Engineering}}}}\ (\bibinfo  {publisher} {John Wiley
  \& Sons, Ltd},\ \bibinfo {address} {Chichester, UK},\ \bibinfo {year}
  {2012})\BibitemShut {NoStop}%
\bibitem [{\citenamefont {Sun}\ \emph {et~al.}(2023)\citenamefont {Sun},
  \citenamefont {Zhang}, \citenamefont {Liu}, \citenamefont {Fan},
  \citenamefont {Li},\ and\ \citenamefont {Wei}}]{sun_research_2023}%
  \BibitemOpen
  \bibfield  {author} {\bibinfo {author} {\bibfnamefont {Y.~Q.}\ \bibnamefont
  {Sun}}, \bibinfo {author} {\bibfnamefont {Y.~P.}\ \bibnamefont {Zhang}},
  \bibinfo {author} {\bibfnamefont {Q.~W.}\ \bibnamefont {Liu}}, \bibinfo
  {author} {\bibfnamefont {Z.~Z.}\ \bibnamefont {Fan}}, \bibinfo {author}
  {\bibfnamefont {N.}~\bibnamefont {Li}}, \ and\ \bibinfo {author}
  {\bibfnamefont {A.~Q.}\ \bibnamefont {Wei}},\ }\href {\doibase
  10.1134/S0965544123080029} {\bibfield  {journal} {\bibinfo  {journal} {Pet.
  Chem.}\ }\textbf {\bibinfo {volume} {63}},\ \bibinfo {pages} {1119} (\bibinfo
  {year} {2023})}\BibitemShut {NoStop}%
\bibitem [{\citenamefont {Chaudhry}\ \emph {et~al.}(2024)\citenamefont
  {Chaudhry}, \citenamefont {Muneer}, \citenamefont {Lashari}, \citenamefont
  {Hashmet}, \citenamefont {Osei-Bonsu}, \citenamefont {Abdala},\ and\
  \citenamefont {Rabbani}}]{chaudhry_recent_2024}%
  \BibitemOpen
  \bibfield  {author} {\bibinfo {author} {\bibfnamefont {A.~U.}\ \bibnamefont
  {Chaudhry}}, \bibinfo {author} {\bibfnamefont {R.}~\bibnamefont {Muneer}},
  \bibinfo {author} {\bibfnamefont {Z.~A.}\ \bibnamefont {Lashari}}, \bibinfo
  {author} {\bibfnamefont {M.~R.}\ \bibnamefont {Hashmet}}, \bibinfo {author}
  {\bibfnamefont {K.}~\bibnamefont {Osei-Bonsu}}, \bibinfo {author}
  {\bibfnamefont {A.}~\bibnamefont {Abdala}}, \ and\ \bibinfo {author}
  {\bibfnamefont {H.~S.}\ \bibnamefont {Rabbani}},\ }\href {\doibase
  10.1016/j.molliq.2024.125209} {\bibfield  {journal} {\bibinfo  {journal}
  {Journal of Molecular Liquids}\ }\textbf {\bibinfo {volume} {407}},\ \bibinfo
  {pages} {125209} (\bibinfo {year} {2024})}\BibitemShut {NoStop}%
\bibitem [{\citenamefont {Harshini}\ \emph {et~al.}(2024)\citenamefont
  {Harshini}, \citenamefont {P.g}, \citenamefont {Kumari},\ and\ \citenamefont
  {Zhang}}]{harshini_innovative_2024}%
  \BibitemOpen
  \bibfield  {author} {\bibinfo {author} {\bibfnamefont {R.~D. G.~F.}\
  \bibnamefont {Harshini}}, \bibinfo {author} {\bibfnamefont {R.}~\bibnamefont
  {P.g}}, \bibinfo {author} {\bibfnamefont {W.~G.~P.}\ \bibnamefont {Kumari}},
  \ and\ \bibinfo {author} {\bibfnamefont {D.~C.}\ \bibnamefont {Zhang}},\
  }\href {\doibase 10.1016/j.geoen.2024.213091} {\bibfield  {journal} {\bibinfo
   {journal} {Geoenergy Science and Engineering}\ }\textbf {\bibinfo {volume}
  {241}},\ \bibinfo {pages} {213091} (\bibinfo {year} {2024})}\BibitemShut
  {NoStop}%
\bibitem [{\citenamefont {Weaire}\ and\ \citenamefont
  {Hutzler}(2000)}]{weaire}%
  \BibitemOpen
  \bibfield  {author} {\bibinfo {author} {\bibfnamefont {D.}~\bibnamefont
  {Weaire}}\ and\ \bibinfo {author} {\bibfnamefont {S.}~\bibnamefont
  {Hutzler}},\ }\href@noop {} {\emph {\bibinfo {title} {The physics of
  foams}}}\ (\bibinfo  {publisher} {Oxford Univ. Press},\ \bibinfo {address}
  {Oxford},\ \bibinfo {year} {2000})\BibitemShut {NoStop}%
\bibitem [{\citenamefont {Petit}\ \emph {et~al.}(2015)\citenamefont {Petit},
  \citenamefont {Seiwert}, \citenamefont {Cantat},\ and\ \citenamefont
  {Biance}}]{petit2015generation}%
  \BibitemOpen
  \bibfield  {author} {\bibinfo {author} {\bibfnamefont {P.}~\bibnamefont
  {Petit}}, \bibinfo {author} {\bibfnamefont {J.}~\bibnamefont {Seiwert}},
  \bibinfo {author} {\bibfnamefont {I.}~\bibnamefont {Cantat}}, \ and\ \bibinfo
  {author} {\bibfnamefont {A.-L.}\ \bibnamefont {Biance}},\ }\href
  {https://doi.org/10.1017/jfm.2014.662} {\bibfield  {journal} {\bibinfo
  {journal} {Journal of Fluid Mechanics}\ }\textbf {\bibinfo {volume} {763}},\
  \bibinfo {pages} {286} (\bibinfo {year} {2015})}\BibitemShut {NoStop}%
\bibitem [{\citenamefont {Saint-Jalmes}\ and\ \citenamefont
  {Trégouët}(2023)}]{saint-jalmes_foam_2023}%
  \BibitemOpen
  \bibfield  {author} {\bibinfo {author} {\bibfnamefont {A.}~\bibnamefont
  {Saint-Jalmes}}\ and\ \bibinfo {author} {\bibfnamefont {C.}~\bibnamefont
  {Trégouët}},\ }\href {\doibase 10.1039/D2SM01618D} {\bibfield  {journal}
  {\bibinfo  {journal} {Soft Matter}\ }\textbf {\bibinfo {volume} {19}},\
  \bibinfo {pages} {2090} (\bibinfo {year} {2023})}\BibitemShut {NoStop}%
\bibitem [{\citenamefont {Cohen-Addad}\ and\ \citenamefont
  {H\"ohler}(2001)}]{cohenaddad01}%
  \BibitemOpen
  \bibfield  {author} {\bibinfo {author} {\bibfnamefont {S.}~\bibnamefont
  {Cohen-Addad}}\ and\ \bibinfo {author} {\bibfnamefont {R.}~\bibnamefont
  {H\"ohler}},\ }\href {https://doi.org/10.1103/PhysRevLett.86.4700} {\bibfield
   {journal} {\bibinfo  {journal} {Phys. Rev. Lett}\ }\textbf {\bibinfo
  {volume} {86}},\ \bibinfo {pages} {4700} (\bibinfo {year}
  {2001})}\BibitemShut {NoStop}%
\bibitem [{\citenamefont {Gopal}\ and\ \citenamefont {Durian}(1995)}]{gopal95}%
  \BibitemOpen
  \bibfield  {author} {\bibinfo {author} {\bibfnamefont {A.~D.}\ \bibnamefont
  {Gopal}}\ and\ \bibinfo {author} {\bibfnamefont {D.~J.}\ \bibnamefont
  {Durian}},\ }\href {https://doi.org/10.1103/PhysRevLett.75.2610} {\bibfield
  {journal} {\bibinfo  {journal} {Phys. Rev. Lett.}\ }\textbf {\bibinfo
  {volume} {75}},\ \bibinfo {pages} {2610} (\bibinfo {year}
  {1995})}\BibitemShut {NoStop}%
\bibitem [{\citenamefont {Chatzigiannakis}\ \emph {et~al.}(2021)\citenamefont
  {Chatzigiannakis}, \citenamefont {Jaensson},\ and\ \citenamefont
  {Vermant}}]{chatzigiannakis_thin_2021}%
  \BibitemOpen
  \bibfield  {author} {\bibinfo {author} {\bibfnamefont {E.}~\bibnamefont
  {Chatzigiannakis}}, \bibinfo {author} {\bibfnamefont {N.}~\bibnamefont
  {Jaensson}}, \ and\ \bibinfo {author} {\bibfnamefont {J.}~\bibnamefont
  {Vermant}},\ }\href {\doibase 10.1016/j.cocis.2021.101441} {\bibfield
  {journal} {\bibinfo  {journal} {Current Opinion in Colloid \& Interface
  Science}\ }\textbf {\bibinfo {volume} {53}},\ \bibinfo {pages} {101441}
  (\bibinfo {year} {2021})}\BibitemShut {NoStop}%
\bibitem [{\citenamefont {Mysels}\ \emph {et~al.}(1959)\citenamefont {Mysels},
  \citenamefont {Frankel},\ and\ \citenamefont {Shinoda}}]{mysels1959soap}%
  \BibitemOpen
  \bibfield  {author} {\bibinfo {author} {\bibfnamefont {K.~J.}\ \bibnamefont
  {Mysels}}, \bibinfo {author} {\bibfnamefont {S.}~\bibnamefont {Frankel}}, \
  and\ \bibinfo {author} {\bibfnamefont {K.}~\bibnamefont {Shinoda}},\
  }\href@noop {} {\emph {\bibinfo {title} {Soap films: studies of their
  thinning and a bibliography}}}\ (\bibinfo  {publisher} {Pergamon press},\
  \bibinfo {year} {1959})\BibitemShut {NoStop}%
\bibitem [{\citenamefont {Lhuissier}\ and\ \citenamefont
  {Villermaux}(2012)}]{lhuissier2012bursting}%
  \BibitemOpen
  \bibfield  {author} {\bibinfo {author} {\bibfnamefont {H.}~\bibnamefont
  {Lhuissier}}\ and\ \bibinfo {author} {\bibfnamefont {E.}~\bibnamefont
  {Villermaux}},\ }\href
  {https://www.cambridge.org/core/journals/journal-of-fluid-mechanics/article/bursting-bubble-aerosols/33D700585872DF6B9A50030FD7FCD0E1}
  {\bibfield  {journal} {\bibinfo  {journal} {Journal of Fluid Mechanics}\
  }\textbf {\bibinfo {volume} {696}},\ \bibinfo {pages} {5} (\bibinfo {year}
  {2012})}\BibitemShut {NoStop}%
\bibitem [{\citenamefont {Gros}\ \emph {et~al.}(2021)\citenamefont {Gros},
  \citenamefont {Bussonni\`ere}, \citenamefont {Nath},\ and\ \citenamefont
  {Cantat}}]{gros2021marginal}%
  \BibitemOpen
  \bibfield  {author} {\bibinfo {author} {\bibfnamefont {A.}~\bibnamefont
  {Gros}}, \bibinfo {author} {\bibfnamefont {A.}~\bibnamefont {Bussonni\`ere}},
  \bibinfo {author} {\bibfnamefont {S.}~\bibnamefont {Nath}}, \ and\ \bibinfo
  {author} {\bibfnamefont {I.}~\bibnamefont {Cantat}},\ }\href {\doibase
  10.1103/PhysRevFluids.6.024004} {\bibfield  {journal} {\bibinfo  {journal}
  {Phys. Rev. Fluids}\ }\textbf {\bibinfo {volume} {6}},\ \bibinfo {pages}
  {024004} (\bibinfo {year} {2021})}\BibitemShut {NoStop}%
\bibitem [{\citenamefont {Cohen}\ \emph {et~al.}(2014)\citenamefont {Cohen},
  \citenamefont {Fraysse}, \citenamefont {Rajchenbach}, \citenamefont
  {Argentina}, \citenamefont {Bouret},\ and\ \citenamefont
  {Raufaste}}]{cohen2014inertial}%
  \BibitemOpen
  \bibfield  {author} {\bibinfo {author} {\bibfnamefont {A.}~\bibnamefont
  {Cohen}}, \bibinfo {author} {\bibfnamefont {N.}~\bibnamefont {Fraysse}},
  \bibinfo {author} {\bibfnamefont {J.}~\bibnamefont {Rajchenbach}}, \bibinfo
  {author} {\bibfnamefont {M.}~\bibnamefont {Argentina}}, \bibinfo {author}
  {\bibfnamefont {Y.}~\bibnamefont {Bouret}}, \ and\ \bibinfo {author}
  {\bibfnamefont {C.}~\bibnamefont {Raufaste}},\ }\href
  {https://doi.org/10.1103/PhysRevLett.112.218303} {\bibfield  {journal}
  {\bibinfo  {journal} {Physical Review Letters}\ }\textbf {\bibinfo {volume}
  {112}},\ \bibinfo {pages} {218303} (\bibinfo {year} {2014})}\BibitemShut
  {NoStop}%
\bibitem [{\citenamefont {Argentina}\ \emph {et~al.}(2015)\citenamefont
  {Argentina}, \citenamefont {Cohen}, \citenamefont {Bouret}, \citenamefont
  {Fraysse},\ and\ \citenamefont {Raufaste}}]{argentina2015one}%
  \BibitemOpen
  \bibfield  {author} {\bibinfo {author} {\bibfnamefont {M.}~\bibnamefont
  {Argentina}}, \bibinfo {author} {\bibfnamefont {A.}~\bibnamefont {Cohen}},
  \bibinfo {author} {\bibfnamefont {Y.}~\bibnamefont {Bouret}}, \bibinfo
  {author} {\bibfnamefont {N.}~\bibnamefont {Fraysse}}, \ and\ \bibinfo
  {author} {\bibfnamefont {C.}~\bibnamefont {Raufaste}},\ }\href
  {https://doi.org/10.1017/jfm.2014.717} {\bibfield  {journal} {\bibinfo
  {journal} {Journal of Fluid Mechanics}\ }\textbf {\bibinfo {volume} {765}},\
  \bibinfo {pages} {1} (\bibinfo {year} {2015})}\BibitemShut {NoStop}%
\bibitem [{\citenamefont {Cohen}\ \emph {et~al.}(2015)\citenamefont {Cohen},
  \citenamefont {Fraysse},\ and\ \citenamefont {Raufaste}}]{cohen2015drop}%
  \BibitemOpen
  \bibfield  {author} {\bibinfo {author} {\bibfnamefont {A.}~\bibnamefont
  {Cohen}}, \bibinfo {author} {\bibfnamefont {N.}~\bibnamefont {Fraysse}}, \
  and\ \bibinfo {author} {\bibfnamefont {C.}~\bibnamefont {Raufaste}},\ }\href
  {https://doi.org/10.1103/PhysRevE.91.053008} {\bibfield  {journal} {\bibinfo
  {journal} {Physical Review E}\ }\textbf {\bibinfo {volume} {91}},\ \bibinfo
  {pages} {053008} (\bibinfo {year} {2015})}\BibitemShut {NoStop}%
\bibitem [{\citenamefont {Koehler}\ \emph {et~al.}(2000)\citenamefont
  {Koehler}, \citenamefont {Hilgenfeldt},\ and\ \citenamefont
  {Stone}}]{koehler_generalized_2000}%
  \BibitemOpen
  \bibfield  {author} {\bibinfo {author} {\bibfnamefont {S.~A.}\ \bibnamefont
  {Koehler}}, \bibinfo {author} {\bibfnamefont {S.}~\bibnamefont
  {Hilgenfeldt}}, \ and\ \bibinfo {author} {\bibfnamefont {H.~A.}\ \bibnamefont
  {Stone}},\ }\href {\doibase 10.1021/la9913147} {\bibfield  {journal}
  {\bibinfo  {journal} {Langmuir}\ }\textbf {\bibinfo {volume} {16}},\ \bibinfo
  {pages} {6327} (\bibinfo {year} {2000})}\BibitemShut {NoStop}%
\bibitem [{\citenamefont {Saint-Jalmes}(2006)}]{saint-jalmes_physical_2006}%
  \BibitemOpen
  \bibfield  {author} {\bibinfo {author} {\bibfnamefont {A.}~\bibnamefont
  {Saint-Jalmes}},\ }\href {\doibase 10.1039/B606780H} {\bibfield  {journal}
  {\bibinfo  {journal} {Soft Matter}\ }\textbf {\bibinfo {volume} {2}},\
  \bibinfo {pages} {836} (\bibinfo {year} {2006})}\BibitemShut {NoStop}%
\bibitem [{Sup()}]{SuppMat}%
  \BibitemOpen
  \href@noop {} {}\bibinfo {note} {See Supplemental Material at [URL will be
  inserted by the publisher] for movies.}\BibitemShut {Stop}%
\bibitem [{\citenamefont {Adami}\ and\ \citenamefont
  {Caps}(2014)}]{adami2014capillary}%
  \BibitemOpen
  \bibfield  {author} {\bibinfo {author} {\bibfnamefont {N.}~\bibnamefont
  {Adami}}\ and\ \bibinfo {author} {\bibfnamefont {H.}~\bibnamefont {Caps}},\
  }\href {https://iopscience.iop.org/article/10.1209/0295-5075/106/46001/meta}
  {\bibfield  {journal} {\bibinfo  {journal} {EPL (Europhysics Letters)}\
  }\textbf {\bibinfo {volume} {106}},\ \bibinfo {pages} {46001} (\bibinfo
  {year} {2014})}\BibitemShut {NoStop}%
\bibitem [{\citenamefont {Plateau}(1873)}]{plateau1873}%
  \BibitemOpen
  \bibfield  {author} {\bibinfo {author} {\bibfnamefont {J.}~\bibnamefont
  {Plateau}},\ }\href@noop {} {\emph {\bibinfo {title} {Statique
  exp\'erimentale et th\'eorique des liquides soumis aux seules forces
  mol\'eculaires}}}\ (\bibinfo  {publisher} {Gauthier-Villars, Paris, France},\
  \bibinfo {year} {1873})\BibitemShut {NoStop}%
\bibitem [{\citenamefont {Couder}\ \emph {et~al.}(1989)\citenamefont {Couder},
  \citenamefont {Chomaz},\ and\ \citenamefont
  {Rabaud}}]{couder1989hydrodynamics}%
  \BibitemOpen
  \bibfield  {author} {\bibinfo {author} {\bibfnamefont {Y.}~\bibnamefont
  {Couder}}, \bibinfo {author} {\bibfnamefont {J.}~\bibnamefont {Chomaz}}, \
  and\ \bibinfo {author} {\bibfnamefont {M.}~\bibnamefont {Rabaud}},\ }\href
  {https://www.sciencedirect.com/science/article/pii/0167278989901449}
  {\bibfield  {journal} {\bibinfo  {journal} {Physica D: Nonlinear Phenomena}\
  }\textbf {\bibinfo {volume} {37}},\ \bibinfo {pages} {384} (\bibinfo {year}
  {1989})}\BibitemShut {NoStop}%
\bibitem [{\citenamefont {Golemanov}\ \emph {et~al.}(2008)\citenamefont
  {Golemanov}, \citenamefont {Denkov}, \citenamefont {Tcholakova},
  \citenamefont {Vethamuthu},\ and\ \citenamefont
  {Lips}}]{golemanov2008surfactant}%
  \BibitemOpen
  \bibfield  {author} {\bibinfo {author} {\bibfnamefont {K.}~\bibnamefont
  {Golemanov}}, \bibinfo {author} {\bibfnamefont {N.}~\bibnamefont {Denkov}},
  \bibinfo {author} {\bibfnamefont {S.}~\bibnamefont {Tcholakova}}, \bibinfo
  {author} {\bibfnamefont {M.}~\bibnamefont {Vethamuthu}}, \ and\ \bibinfo
  {author} {\bibfnamefont {A.}~\bibnamefont {Lips}},\ }\href
  {https://doi.org/10.1021/la8015386} {\bibfield  {journal} {\bibinfo
  {journal} {Langmuir}\ }\textbf {\bibinfo {volume} {24}},\ \bibinfo {pages}
  {9956} (\bibinfo {year} {2008})}\BibitemShut {NoStop}%
\bibitem [{\citenamefont {Monier}\ \emph {et~al.}(2024)\citenamefont {Monier},
  \citenamefont {Gauci}, \citenamefont {Claudet}, \citenamefont {Celestini},
  \citenamefont {Brouzet},\ and\ \citenamefont
  {Raufaste}}]{monier_self-similar_2024}%
  \BibitemOpen
  \bibfield  {author} {\bibinfo {author} {\bibfnamefont {A.}~\bibnamefont
  {Monier}}, \bibinfo {author} {\bibfnamefont {F.-X.}\ \bibnamefont {Gauci}},
  \bibinfo {author} {\bibfnamefont {C.}~\bibnamefont {Claudet}}, \bibinfo
  {author} {\bibfnamefont {F.}~\bibnamefont {Celestini}}, \bibinfo {author}
  {\bibfnamefont {C.}~\bibnamefont {Brouzet}}, \ and\ \bibinfo {author}
  {\bibfnamefont {C.}~\bibnamefont {Raufaste}},\ }\href {\doibase
  10.1103/PhysRevFluids.9.124001} {\bibfield  {journal} {\bibinfo  {journal}
  {Phys. Rev. Fluids}\ }\textbf {\bibinfo {volume} {9}},\ \bibinfo {pages}
  {124001} (\bibinfo {year} {2024})}\BibitemShut {NoStop}%
\end{thebibliography}%

\newpage
\appendix

\section{END MATTER}

\section{Linear flux for the liquid exchange between meniscus and soap film}

\begin{figure}[h!]
\centering
\includegraphics[width=0.45\textwidth]{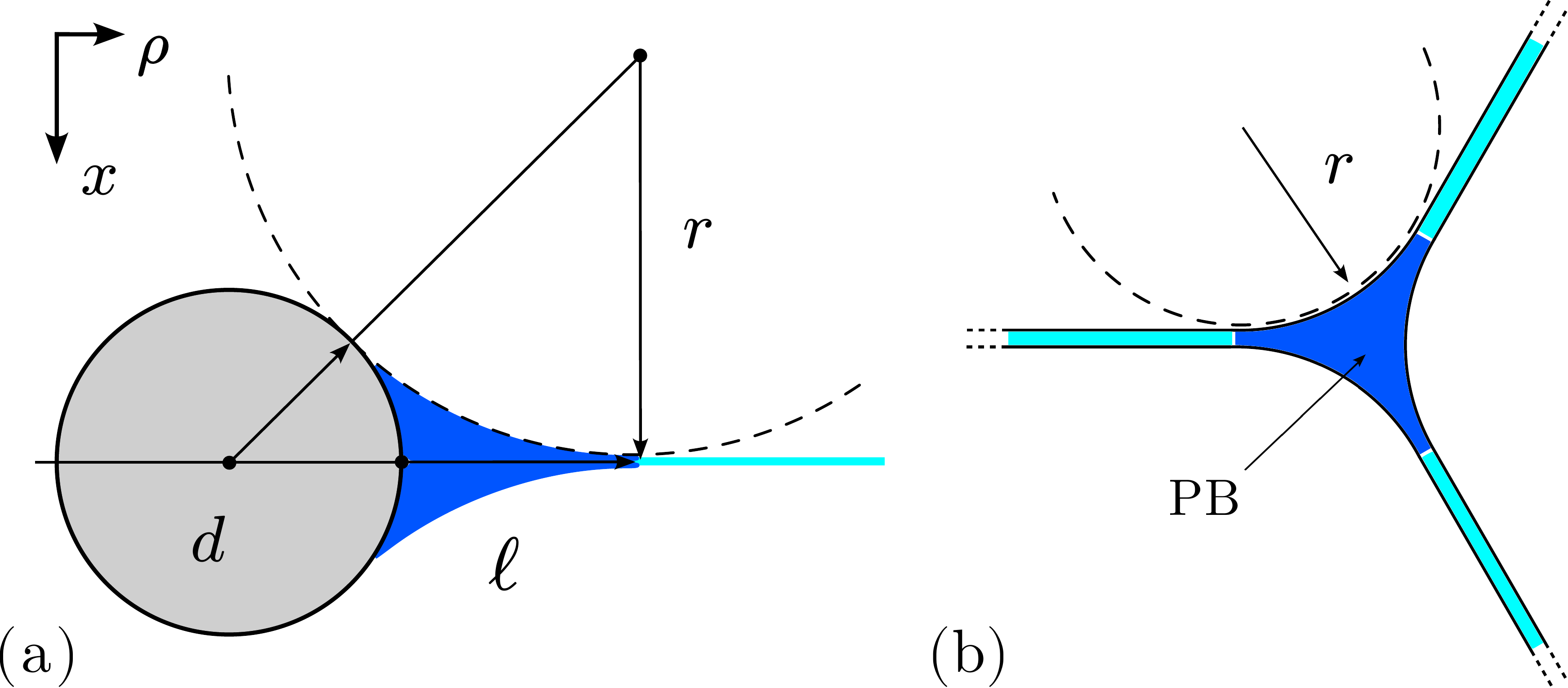}
\caption{
(a)~Sketch of the cross-section of a meniscus (blue) attached to the ring (gray) and the soap film (cyan). The ring cross-section diameter~$d$, the meniscus radius~$r$, and its extent ~$\ell$ are defined. The film thickness is assumed to be small compared to these lengths.
(b)~Sketch of the cross-section of a Plateau border (PB) in blue, at the intersection between three soap films shown in cyan. The curvature radius of the PB is $r$.
}
\label{Fig:relation_btw-r_m-and-l}
\end{figure}

The capillary suction of liquid from the film to the meniscus results from a difference in Laplace pressure, given by $\gamma/r$, between a flat film of thickness $h$ and a concave meniscus of radius $r$, where $\gamma$ is the surface tension (Fig.~\ref{Fig:relation_btw-r_m-and-l}). The linear flux (per unit length of contact between the film and meniscus) can be estimated by considering a Poiseuille flow: in the meniscus, the Laplace pressure scales as $\gamma/r$ and decreases over a characteristic length $w = \sqrt{h r}$ to approach zero in the film~\cite{mysels1959soap}. The typical suction speed $v_s$ can be found from the Stokes equation, $\gamma/(r w) \sim \eta v_s/h^2$, where $\eta$ is the dynamic viscosity. Consequently, the linear flux scales as $h v_s \sim (h \gamma/\eta) (h/r)^{3/2}$~\cite{lhuissier2012bursting}. Although the marginal regeneration process involves more complex flow dynamics, it follows this fundamental scaling~\cite{gros2021marginal}.

\section{Surfactant solutions and thickness measurements}\label{App:SurfactantSolutions}

\begin{figure}[h!]
\centering
\includegraphics[width=0.45\textwidth]{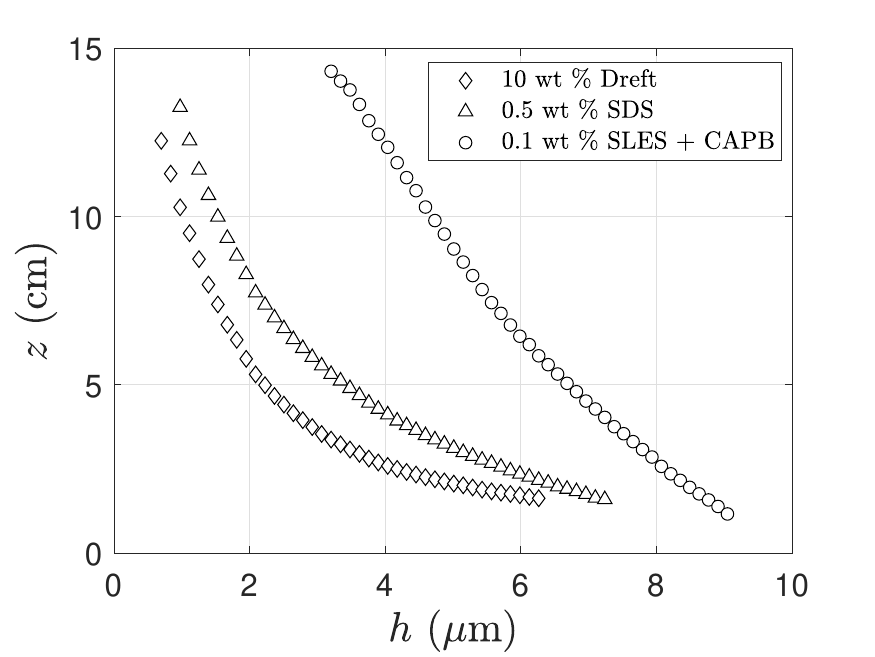}
\caption{
Examples of steady-state thickness profiles of soap films obtained with the three surfactant solutions: SLES-CAPB, SDS, and Dreft, at flow rates of 0.6, 1.0, and 1.1~mL.min$^{-1}$, respectively.
}
\label{Fig:ThicknessProfiles}
\end{figure}

Three surfactant solutions were used, all prepared with deionized water.
The first is a SLES-CAPB solution obtained following the protocol of Golemanov \textit{et al.}~\cite{golemanov2008surfactant}. It consists of a mixture of the two surfactants in a 2/3 SLES and 1/3 CAPB ratio, with a concentration of $0.1$~wt\% in water. The surface tension and viscosity were measured to be $24$~mN/m and $1$~mPa.s, respectively.
The second is a 85/15 water-glycerol mixture by weight, with sodium dodecyl sulfate (SDS) at a concentration of $5.6$~g/L~\cite{monier_self-similar_2024}. The surface tension and viscosity were measured to be $35$~mN/m and $1.5$~mPa.s, respectively.
Finally, we used a commercial surfactant, Dreft (Procter \& Gamble), at a $10$~wt\% concentration in water \cite{cohen2014inertial}. The surface tension and viscosity were measured to be $26$~mN/m and $1$~mPa·s, respectively.
The density of each solution is approximately equivalent to that of water, leading to $\lambda_c = 1.56$~mm for the SLES-CAPB solution, $\lambda_c = 1.89$~mm for the SDS solution and $\lambda_c = 1.63$~mm for the Dreft solution. 
Thickness profiles were obtained using interference fringes supplemented by a spectrometer to have a reference value at a given position (see examples in Fig.~\ref{Fig:ThicknessProfiles}).

\section{Geometrical considerations}

Assuming that the film thickness is very small with respect to the other lengths in the problem and that the meniscus wets the ring, there is a simple geometrical relation between the radius of curvature~$r$ and extent~$\ell$ of the meniscus, and the minor diameter of the ring~$d$, defined in Fig.~\ref{Fig:relation_btw-r_m-and-l}(a). This relation is 
\begin{equation}\label{Eq:rm_full}
r = \ell \left(1+\frac{\ell}{d}\right).
\end{equation}
The meniscus area~$A$, representing the area of the blue region in Fig.~\ref{Fig:relation_btw-r_m-and-l}(a), is given by
\begin{multline}\label{Eq:A_full}
A= r \left(\frac{d}{2} + \ell \right) - \arcsin\left(\frac{r}{r + \frac{d}{2}}\right) \frac{d^2}{4} \\
- \arccos\left(\frac{r}{r + \frac{d}{2}}\right) r^2 .
\end{multline}
In addition, if one considers the small minor diameter limit as in the main body of the text, such as $d \ll~\ell~\ll~r$, Eqs.~\eqref{Eq:rm_full} and~\eqref{Eq:A_full} lead to $r \simeq \ell^2 /d$ and $A \simeq \ell \, d/3$, respectively. In the other limit discussed at the end of the paper where $d \gg r$, one has $r \simeq \ell$ and $A\simeq r^2 (2-\pi/2)$.

In the case of a meniscus or Plateau border formed at the intersection of three soap films (Fig.~\ref{Fig:relation_btw-r_m-and-l}(b)), the only relevant length scale is $r$. Consequently, we expect $\ell \propto r$ and $A\propto r^2$, similar to the case of a meniscus in contact with an object in the limit  $d \gg r$.

\section{Equations in the limit of large minor diameter}

In the limit of large minor diameter $d \gg r$, $r\simeq l$ and $A\simeq r^2 (2-\pi/2)$. Implementing these approximations into Eq.~\eqref{Eq:fluxbalance} leads to
\begin{equation}\label{Eq:dimensionequation_approx2}
    r^{3/2} \frac{\mathrm{d}}{\mathrm{d}\theta}\left(r^{4} \sin\theta - \frac{2 \lambda_c^2 r^2}{D}\frac{\mathrm{d}r}{\mathrm{d}\theta}\right) = k \lambda_c^{2}D\,h^{5/2},
\end{equation}
which is independent of the minor diameter~$d$. Setting the gravito-exchange length in this limit to $r_g=k^{2/11}\lambda_c^{4/11}D^{2/11}h^{5/11}$, we obtain a dimensionless equation for $\tilde{r}=r/r_s$ as
\begin{equation}\label{Eq:dimensionlessequation_approx2}
    \tilde{r}^{3/2} \frac{\mathrm{d}}{\mathrm{d}\theta}\left(\tilde{r}^{4} \sin\theta - 2\tilde{r}^{2} \frac{\mathrm{d}\tilde{r}}{\mathrm{d}\theta}\right) = (r_g/r_s)^{11/2} ,
\end{equation}
where $r_g/r_s=k^{2/11}\lambda_c^{-18/11}D^{13/11}h^{5/11}$.
Note that the constant~$k$ is still expected of order~$1$, but is in principle different to the value found in the small minor diameter approximation $d \ll r$. The hydrostatic solution is the same as the one in the other limit, but the solution for $r_g/r_s\neq 0$ is different. However, since Eq.~\eqref{Eq:dimensionlessequation_approx2} is similar to Eq.~\eqref{Eq:dimensionlessequation_all}, one can expect the same type of behavior in this other limit. In particular, when $r_g/r_s$ is large, the radius of curvature is expected to start at $r_0\approx r_g$ and to follow the hydrostatic case when $\theta$ approaches~$\pi$.

\section{Determination of $k$}
In the flowing regime and $d \ll r$ approximation, we expect $r_0 =k^{1/3} \lambda_c^{2/3}D^{1/3}d^{-5/6}h^{5/6}$. To determine the unknown constant $k$, we performed a proportionality fit between the measured values of $r_0$ and $\lambda_c^{2/3}D^{1/3}d^{-5/6}h^{5/6}$, where the coefficient of proportionality provides $k^{1/3}$. To ensure the analysis is restricted to the flowing regime and $d \ll r$ approximation, we considered only data where $r_0 \geq 5\,r_s$ and the two smallest minor diameters, $d=14.5$ and $52~\mu$m.

\end{document}